# The "Seen but Unnoticed" Vocabulary of Natural Touch: Revolutionizing Direct Interaction with Our Devices and One Another

Ken Hinckley

Microsoft Research, kenh@microsoft.com

This UIST Vision argues that "touch" input and interaction remains in its infancy when viewed in context of the *seen but unnoticed* vocabulary of natural human behaviors, activity, and environments that surround direct interaction with displays. Unlike status-quo touch interaction—a shadowplay of fingers on a single screen—I argue that our perspective of direct interaction should encompass the full rich context of individual use (whether via touch, sensors, or in combination with other modalities), as well as collaborative activity where people are engaged in local (co-located), remote (tele-present), and hybrid work. We can further view touch through the lens of the "Society of Devices," where each person's activities span many complementary, oft-distinct devices that offer the right task affordance (input modality, screen size, aspect ratio, or simply a distinct surface with dedicated purpose) at the right place and time. While many hints of this vision already exist (e.g. [1, 4, 13, 17, 48]), I speculate that a comprehensive program of research to systematically inventory, sense, and design interactions around such human behaviors and activities—and that fully embrace touch as a multi-modal, multi-sensor, multi-user, and multi-device construct—could revolutionize both individual and collaborative interaction with technology.

## 1 INTRODUCTION

Within the past ~15 years, touch input—seemingly well-understood [23]—has become the predominate means of interacting with devices such as smartphones, tablets, and large displays. Yet I argue that much remains unknown—in the form of a *seen but unnoticed* [14] vocabulary of natural touch—with huge untapped potential.

For example, touchscreens remain largely ignorant of the human activity, manual behavior, and context-of-use beyond the moment of finger-contact with the screen itself (e.g. [7, 21, 47, 52]). In a sense, status-quo interactions are trapped in a flatland of touch, while systems remain oblivious to the vibrant world of human behavior, activity, and movement that surrounds them. Such behaviors include how people hold, use, and manipulate the devices themselves [12, 45], as well as the "body language" of how they position their bodies, limbs, and hands while engaging with displays [29] across a range of scales [48]—from smartphones, tablets, and adjustable drafting-tables all the way up to large electronic whiteboards. Indeed, these behaviors are influenced by the presence of other persons (due to sociological factors of inter-personal space, known as *proxemics* [9, 16]), as well as interaction across multiple devices in ad-hoc ensembles [3, 34].

Surfacing these seen-but-unnoticed behaviors entails a holistic re-examination of direct-touch interaction with devices [6, 11, 31, 50]. That each form-factor recruits different limb segments and demands different physical postures and movements is obvious—*once seen and noticed.* We adjust the posture of our human bodies in reference to the devices we use, and traverse our work-spaces throughout the day—moving proximal to some devices [24] even as others fade to the periphery of our attention [44]. And at a more fine-grained scale, our arms and hands move in continuous analog fashion before they touch a screen, and continue to do so after [7, 18, 21, 47, 51]. If devices are mobile or semi-fixed, people grasp [15, 27, 43], adjust, and re-position their displays to afford a particular use [20, 35, 40]—or to share the screen with others nearby, as in collaborative micro-mobility of physical artifacts [32]. Yet this broad diversity of behaviors surrounding the moment of "touch" itself have rarely been characterized in a scientific manner, much less sensed and translated into practice in a manner that could transform people's everyday experiences with mass-market devices.

To unpack and fully realize this vocabulary of naturally-occurring gestures hiding in plain sight, the field must (1) rigorously observe, analyze, and document these behaviors in a wide diversity of the population, including those with mobility, vision, or motor disabilities [36, 37]; (2) develop pragmatic in-device sensors capable of detecting these phenomena robustly; and (3) show how their detection transforms and simplifies user experiences via gestures, sensing techniques, and scenarios-of-use beyond current touchscreen technology.

The potential goes beyond the touchscreen of any particular device. Properly realizing this vision requires extended sensing of not only the user (i.e., their hands, body, and limbs) relative to the screen, but also of devices with respect to one another [2]. These concepts encompass device-centric constructs, such as the "posture" embodied by the device itself [19, 40, 52], as well as human- or group-centric constructs, such as the spatial relationship of multiple collaborating users. The latter includes proxemic notions such as the spacing of people between devices and one another [16, 42], as well as relative body orientation and clustering of persons in focused encounters ("f-formations") [8, 25, 26, 33]. But very little is known about "device proxemics" and how they influence people's social (socio-spatial) behavior [10, 28, 30, 32] or perception of peripersonal space—i.e. the nearby space within arm's reach, which can be influenced by tools (or devices?) held in-hand [39].

Indeed, the scientific literature lacks a precise vocabulary to distinguish between these multiple frames-of-reference in touch at different scales, from ego-centric to exo-centric, including hand-, device-, user-, group-, and room- or environment-centric perspectives. Ultimately these all must be reconciled in order for human gestures and activity to be ascribed the correct (often implicit, or tacitly understood) meaning.

For example, consider touch in the context of multi-user interaction on large displays. If two or more users stand at a large display and the system registers a touch event, who touched [46] the display? If another user approaches, does that change the semantics of the interaction? What of users in the audience, at a distance, who point or wave at the display? Are such gestures intended for the device, or merely the spontaneous gesticulations that accompany speech? And even if the answer is the latter, are these perhaps contextual cues that could drive subtle background interactions [5] or casual in-air gestures [38] that may enhance meetings, brainstorming, collaboration, presentations, or other shared activities? If so, can they deliver value without being cursed by issues of "Midas Touch" that trigger overeager false positive actions?

My belief is that the potential gains are not incremental, and indeed may open up whole new scenarios, styles of interaction, and vocabularies of gesture. Furthermore, this multi-modal, multi-sensor, multi-user, and multi-device perspective of "touch" [22] goes hand in hand with a technological sea-change that we are currently witnessing. For traditional silicon, Moore's Law is essentially over. Yet, the past decade has witnessed exponential growth in networking and storage price/performance. Hence, in a Computer Science sense these relative performance trends appear poised to drive a radical shift in the way systems (i.e. multi-device distributed systems [4]) are programmed and ultimately experienced by the end-user. For example, what does it mean to have an operating system process (application) that spans multiple, distinct computers? Do the answers here change if the devices involved are owned by the same person, multiple co-located persons, or even multiple remotely located (or telepresent) users and organizations?

Beyond these technical considerations, user productivity clearly shouldn't stop at any particular screen bezel. Not unlike paper-bound office workers of the past [41], modern information workers want to combine and cross-reference information from multiple sources. In the digital world this means multiple devices of diverse form-factors nearby. People also desire to expand screen real estate, to combine displays of different size and aspect ratio, to compose specialized device capabilities such as pen, touch, camera, or microphone inputs, as well as to collaborate in a digitally authentic way with colleagues whether co-located, remote, or hybrid.

I believe such research questions and technology trends will shape the future of HCI. The answers have profound implications for computer science, including cross-device and multi-device user experiences that go far beyond any status-quo notion of touchscreens and interaction via "touch." Direct interaction will extend above, around, between, and across multiple input modalities, devices, contexts, and persons. And if our field gets it right, much of this experience will be driven by a vocabulary of natural touch [49], perhaps still largely seen but unnoticed, yet poised to revolutionize interaction with our devices and—most importantly—our own human potential and collaborative interactions with one another.

## ACKNOWLEDGMENTS

Christian Holz spurred and helped develop these ideas, including proposal for ETH Zurich via Swiss Joint Research Center with Microsoft Research. I also thank the EPIC Research and Surface Fleet teams at Microsoft Research, including Jonathan Goldstein for computer science and distributed systems implications of recent technology trends. Likewise, many of perspectives articulated herein have been shaped by conversations with Nicolai Marquardt, Saul Greenberg, Bill Buxton, Andy Wilson, Abigail Sellen, and other colleagues worldwide.